\documentclass[12pt]{article} 

\newcommand{\be}{\begin{equation}}
\newcommand{\ee}{\end{equation}}
\newcommand{\bear}{\begin{eqnarray}}
\newcommand{\eear}{\end{eqnarray}} \newcommand{\ba}{\begin{array}}
\newcommand{\ea}{\end{array}}
\newcommand{\lae}{\begin{array}{c}\,\sim\vspace{-21pt}\\<
\end{array}}
\newcommand{\gae}{\begin{array}{c}\,\sim\vspace{-21pt}\\>
\end{array}}

 \newcommand{\CQ}{{\cal Q}}
\newcommand{\CU}{{\cal U}} \newcommand{\CD}{{\cal D}}
\newcommand{\CL}{{\cal L}} \newcommand{\CE}{{\cal E}}
 
\newcommand{\CH}{{\cal H}}

\begin{document}

\pagestyle{empty} \begin{titlepage}
\def\thepage {}        

\title{\Large \bf Universal Extra Dimensions and \\ [2mm]
the Muon Magnetic Moment 
\\ [1.3cm]}

\author{\normalsize
\bf \hspace*{-.3cm} Thomas Appelquist and Bogdan A.~Dobrescu
 \\ \\ {\small {\it
\vspace*{-5cm}
Department of Physics, Yale University, New
Haven, CT 06511, USA\thanks{e-mail: thomas.appelquist@yale.edu, \
bogdan.dobrescu@yale.edu}}}\\
 }

\date{ } \maketitle

\vspace*{-7.9cm}
\noindent \makebox[12.7cm][l]{\small \hspace*{-.2cm}
hep-ph/0106140} {\small YCTP-P4-01 } \\
\makebox[11.8cm][l]{\small \hspace*{-.2cm} June 13, 2001 }
{\small } \\

 \vspace*{10.5cm}

  \begin{abstract}
{\small We analyze the muon anomalous magnetic moment in the
context of universal extra dimensions. Our computation shows that
the bound from electroweak data on the size of these
dimensions allows only a small shift in the muon magnetic moment
given by Kaluza-Klein modes of standard model fields. In the
well-motivated case of two universal extra dimensions, additional
contributions arising from physics at scales where the effective
6-dimensional standard model breaks down, given by
dimension-ten operators, have a natural size comparable to
the sensitivity of the muon $(g-2)$ experiment at BNL.}
\end{abstract}

\vfill \end{titlepage}

\baselineskip=18pt \pagestyle{plain} \setcounter{page}{1}

\section{Introduction} \setcounter{equation}{0}

There are good reasons to imagine that {\it all} the standard model
fields propagate in a larger number of spatial dimensions
compactified at a scale $1/R \lae 1$ TeV. This framework could
provide a mechanism for electroweak symmetry breaking
\cite{Arkani-Hamed:2000hv} and supersymmetry breaking
\cite{Barbieri:2001vh}, and it relates the number of fermion
generations to the requirement of anomaly cancellation
\cite{Dobrescu:2001ae}. Recently, it was pointed out that the
compactification radius $R$ of these universal extra dimensions,
can be surprisingly large, as large as $ 1/(300 \; {\rm GeV})$.
The reason is that momentum conservation in extra dimensions
leads to Kaluza-Klein (KK) number conservation and therefore to
the absence of vertices with a single nonzero KK mode. There are
thus no tree-level contributions to the electroweak observables,
and no single KK mode production at colliders. Interestingly, the
tightest bounds on $R$, derived from the experimental constraints
on the $\rho$ parameter \cite{Appelquist:2000nn} and on the $b
\rightarrow s\gamma$ process \cite{Agashe:2001xt}, leave room for
a discovery of KK modes in Run II at the Tevatron.

The higher dimensional standard model
is an effective field theory, valid below some scale $M_s$ in the
multi-TeV range. Its lowest dimension operators correspond
directly to the familiar terms of the 4-dimensional standard
model. Corrections to the leading low energy theory are encoded in
a tower of operators of increasing dimension allowed by the field
content and the symmetries. This effective higher dimensional
theory, after compactification to four dimensions, leads to the
standard model together with two classes of corrections. The first
arises from physics above $1/R$ but below the cutoff $M_s$, and
corresponds to virtual KK modes of the standard model particles.
The second arises from the unknown physics at scales $M_s$ and
above, and is parametrized by the coefficients of the tower of
higher-dimension operators.

In this paper, we extend the considerations of
Ref.~\cite{Appelquist:2000nn} by discussing  the muon anomalous 
magnetic moment in the context of universal extra dimensions. We
compute the one-loop contribution from standard model KK modes in
universal extra dimensions, and find that it is too small to be
seen by the Muon $g-2$ Collaboration \cite{Brown:2001mg}. This
computation confirms the estimate of Agashe, Deshpande and Wu
\cite{Agashe:2001ra}.

We then consider the effects of the higher-dimension operators.
Concentrating on the well motivated case of the (chiral)
6-dimensional standard model, we find that the contributions to
the muon magnetic moment from dimension-ten operators could
naturally be large enough to be experimentally measurable for
typical values of $M_s$, of a few TeV.

In Section 2, we discuss universal extra dimensions in the
framework of effective field theory, 
and display the leptonic sector of this
theory, appropriate for the computation of the muon anomalous 
magnetic moment. In Section 3, we compute at one-loop
level the muon magnetic moment induced by the KK modes
of standard model fields,
and in Section 4 we analyze higher-dimension operators within the
chiral 6-dimensional standard model.
Conclusions and a comparison with the $\rho$ parameter computation
 are contained in Section 5.

\section{Universal extra dimensions}
\setcounter{equation}{0}

The idea of universal extra dimensions is very simple:
{\it all} standard model fields propagate in some extra
spatial dimensions. These universal  dimensions are taken to be
flat, and must be compactified on an orbifold such that the
zero-mode fermions are endowed with 4-dimensional chirality.
The simplest $\delta$-dimensional orbifold of this kind is obtained
by compactifying each pair of extra dimensions on
$T^2/Z_2$ and, for odd $\delta$, the remaining extra dimension
on $S^1/Z_2$, as explicitly shown in \cite{Appelquist:2000nn}.
The ($4+\delta$)-dimensional quarks,
$\CQ_i, \CU_i, \CD_i$, and leptons, $\CL_i, \CE_i$,
($i=1,2,3$ is a generational index)
are decomposed in KK modes such that only one left-
(right-) handed component of each weak doublet (singlet)
is even under the orbifold projection.
For example, the second generation lepton
fields have zero-modes $\CL_2^{(0)} = ({\nu_\mu}_L, \mu_L)$
and $\CE_2^{(0)} = \mu_R$.

The ($4+\delta$)-dimensional Lagrangian looks very similar to
that of the 4-dimensional standard model. There are
kinetic terms for the ($4+\delta$)-dimensional
$SU(3)_C\times SU(2)_W \times U(1)_Y$ gauge fields,
a kinetic term and potential for the
($4+\delta$)-dimensional Higgs doublet, $\CH$, as well as the 
following terms involving quarks and leptons:
\be
\hspace*{-0.71em}
\left(\overline{\CQ}, \overline{\CU},
\overline{\CD},\overline{\CL}, \overline{\CE}\right)
i\Gamma^\alpha D_\alpha \left(\CQ,
\CU, \CD, \CL, \CE \right)
- \left[\overline{\CQ}\left(\hat{\lambda}_\CU  \CU
 i\sigma_2 \CH^* + \hat{\lambda}_\CD \CD \CH\right) +
\overline{\CL} \hat{\lambda}_\CE \CE \CH +
{\rm h.c.} \right]
\label{lagrangian}
\ee
We use the notation $x^{\alpha}$ or $x^{\beta}$
for all space-time coordinates ($\alpha,\beta  = 0,1,...,4+\delta $),
and $x^{\rho}$ or $x^{\tau}$  for the non-compact coordinates
($\rho, \tau = 0,1,2,3$).
$D_\alpha $ are the covariant derivatives
associated with the $SU(3)_C\times SU(2)_W \times U(1)_Y$ group.
$\Gamma^\alpha$ are anti-commuting
$2^{k + 2}\times 2^{k + 2}$ matrices, where $k$ is an integer
such that $\delta = 2k$ or $\delta = 2k+1$.
When $\delta$ is even, the quark and lepton fields may have
$4+\delta$ chirality,
defined by the eigenvalues $\pm 1$ of $\Gamma_{4+\delta}$,
the analogue of $\gamma_5$ in four dimensions.
Anomaly considerations are
discussed in Refs.~\cite{Appelquist:2000nn, Dobrescu:2001ae}.
A summation over a generational index is implicit in
Eq.~(\ref{lagrangian}).
The $(4+\delta)$-dimensional Yukawa couplings,  $\hat{\lambda}_\CU,
 \hat{\lambda}_\CD, \hat{\lambda}_\CE$, are $3\times 3$ matrices
and have mass dimension ${-\delta/2}$.

The standard model operators listed above 
have mass dimension ranging up to $4+ 2\delta$. They are the
lowest dimension operators allowed by gauge symmetry. 
Corrections are described by a tower of higher-dimension
operators, each suppressed by inverse powers of the
multi-TeV scale $M_s$ at which the effective theory breaks down.

In Ref.~\cite{Appelquist:2000nn}, this effective theory was used
to show that, because the KK number is conserved and thus the
contributions to experimental observables arise only from loops,
the bound from the electroweak data on the size of universal
extra dimensions is rather loose. The main constraint comes from
weak-isospin violating effects, encoded in the $\Delta\rho =
\alpha T$ parameter. In the case of a single extra dimension the
electroweak parameters may be computed reliably at one-loop
level using the Lagrangian (\ref{lagrangian}), revealing that the
compactification scale $1/R$ could be as low as 300 GeV. Higher
order corrections, suppressed by inverse powers of $RM_s$, are
small. In the case of two universal extra dimensions, the
contributions of the KK modes to electroweak observables
become logarithmically sensitive to $M_s$, meaning that they are
not reliably computable by relying only on physics below $M_s$. 
A rough
estimate can be made, however, indicating that in this case $1/R$
could be as low as roughly 500 GeV. We return to this discussion
in section 5, after describing our results for the muon $g-2$.

The 4-dimensional Lagrangian is obtained by dimensional
reduction from the $(4+\delta)$-dimensional theory.
The decomposition of the standard model fields in KK modes leads to a
variety of trilinear and quartic interactions.
Contributions to the one-loop muon
anomalous magnetic moment arise
from the leptonic part of Eq.~(\ref{lagrangian}).
The vector-like KK modes associated with the weak-doublet,
${\CL}^{j}_2 = ({\CL}_{\nu_\mu}^{j}, {\CL}_\mu^{ j})$,
and -singlet, ${\CE}_2^{j}$, muon fields have electroweak
symmetric masses $M_j$, with $j \ge 1$. In the case of one
universal extra dimension $M_j = j/R$, while for more dimensions the
KK spectrum is denser.
The zero-mode Higgs doublet aquires a
VEV, breaking the electroweak symmetry, and leading to
mass mixing between the ${\CL}_\mu^{ j}$ and ${\CE}_2^{j}$
KK modes, level by level.
The mixing angle is suppressed by the ratio of the muon mass to
KK mass,
\be \sin\alpha_j \approx \frac{m_\mu}{2M_j} + {\cal
O}\left(m_\mu^3/M_j^3 \right) ~.
\ee

\vspace{-2mm}
The trilinear interaction of the zero-mode muon mass eigenstate, 
$\mu^\prime$,
with the higher KK modes of the $Z$ boson, $Z_\alpha^j$,
and the muon KK mass eigenstates, ${\CL}^{\prime
j}_\mu$ and ${\CE}^{\prime j}_2$, is
described by the following terms in the 4-dimensional Lagrangian:
\bear
&& \hspace*{-1.9em}
\frac{g}{\cos\theta_W} \left\{ \; Z_\rho^j \left[ \;
\overline{\CL}^{\prime j}_\mu \gamma^\rho \left( g_{\mu_L} \cos\alpha_j P_L +
g_{\mu_R} \sin\alpha_j P_R \right)
 - \overline{\CE}^{\prime j}_2
\gamma^\rho \left( g_{\mu_R} \cos\alpha_j P_R + g_{\mu_L} \sin\alpha_j
P_L \right) \right]\mu^\prime
\right.\nonumber \\ [3.2mm]
&& \hspace*{-2.2em} \left.  - i
Z_4^j \left[ \; \overline{\CL}^{\prime j}_\mu \left( g_{\mu_L}
\!\cos\alpha_j P_L - g_{\mu_R} \!\sin\alpha_j P_R \right)
- \overline{\CE}^{\prime j}_2 \left( g_{\mu_R} \!\cos\alpha_j
P_R - g_{\mu_L} \!\sin\alpha_j P_L \right) \right]\mu^\prime \right\}
+ {\rm h.c.} 
 \label{z-coupling}
\eear
As usual, $g$ is the $SU(2)_W$ gauge coupling,
$P_{L,R} = (1\mp \gamma_5)/2$ and
\be
g_{\mu_L} = - \frac{1}{2} + \sin^2\theta_W \; \; , \;  \;  \;
g_{\mu_R} = \sin^2\theta_W ~.
\ee
In Eq.~(\ref{z-coupling}) we have displayed the interactions 
involving the KK modes of only one scalar component of the $Z$. 
For $\delta$ extra dimensions
there are $\delta$ scalars associated with a gauge boson, at each KK level.
For the one-loop computations that we perform in the next section
each scalar KK mode contributes by the same amount, so it is sufficient
to consider the exact form of the interactions involving $Z_4^j$.

Another interaction entering the muon $g-2$ computation is that of
a photon KK mode, $A_{\rho}^j$ and $A_4^j$, with a muon zero-mode
and a muon $j$-mode. It may be obtained from Eq.~(\ref{z-coupling})
by substituting $ Z_{\rho, 4}^j$ with $- A_{\rho, 4}^j
\sin\theta_W\cos\theta_W$ and setting $g_{\mu_L} = g_{\mu_R} =1$.
It is also straightforward to write the interactions of the $W$
scalar KK modes with a muon zero-mode and a muon-neutrino KK mode in the 
weak eigenbasis:
\be
- \frac{i g}{\sqrt{2}} {W_4^+}^j \,  {\overline{\nu}^j_\mu}_R \mu_L
+ {\rm h.c.} 
\ee
The ${W_\rho^+}^j {\overline{\nu}^j_\mu}_L \gamma^\rho
\mu_L$ vertex, involving the $W$ boson KK modes, as well as the
$A_\rho {W_\rho^+}^j {G^-}^j$ and ${G^+}^j
{\overline{\nu}^j_{\mu}}_L \mu_R$ vertices, involving the KK modes of the
charged Goldstone boson eaten by the $W$, are identical with the
standard model ones for the corresponding zero-modes. Finally, the
interactions of the photon zero-mode with the muon or $W$ boson
KK modes are diagonal and determined by the corresponding
electric charge.

\section{$g_\mu - 2$ from KK modes of standard model fields}
\setcounter{equation}{0}

The anomalous magnetic moment of the muon is the coefficient
$a_\mu \equiv g_\mu-2$ in the 4-dimensional momentum space
operator \be - a_\mu \frac{e}{2m_\mu} A_\rho(p^{\rm out} - p^{\rm in})
\overline{\mu}^\prime(p^{\rm out})i \sigma^{\rho\tau} (p_\tau^{\rm out} -
p_\tau^{\rm in}) \mu^\prime(p^{\rm in}). \ee In this section, we compute the
contribution $a_\mu^{\rm KK}$ arising from KK modes associated
with universal extra dimensions.

\begin{figure}[t]
\centering
\begin{picture}(100,130)(200,-1)
\thicklines
\multiput(99,100)(0,20){2}{\qbezier(0,0)(4,5)(0,10)}
\multiput(99,110)(0,20){1}{\qbezier(0,0)(-4,5)(0,10)}
\put(100,100){\line(3, -4){55}}
\put(100,100){\line(-3, -4){55}}
\multiput(55,40)(20, 0){5}{\qbezier(0,0)(5,4)(10,0)}
\multiput(55,40)(20, 0){4}{\qbezier(10,0)(15,-4)(20,0)}
\put(32,65){$\CL^{\prime j}_\mu, \, \CE^{\prime j}_2 $}
\put(80,18){$Z_\rho^j, \, A_\rho^j$}
\multiput(249,100)(0,20){2}{\qbezier(0,0)(4,5)(0,10)}
\multiput(249,110)(0,20){1}{\qbezier(0,0)(-4,5)(0,10)}
\multiput(250,100)(-12,-16){4}{\qbezier(0,0)(-4,-2)(-6,-8)}
\multiput(244,92)(-12,-16){3}{\qbezier(0,0)(0,-4)(-6,-8)}
\put(206,40){\qbezier(0,0)(2,1)(2,4)}
\put(207,40){\line(-3, -4){10}}
\multiput(250,100)(12,-16){4}{\qbezier(0,0)(4,-2)(6,-8)}
\multiput(256,92)(12,-16){3}{\qbezier(0,0)(0,-4)(6,-8)}
\put(294,40){\qbezier(1,0)(-2,1)(-1,4)}
\put(295,40){\line(3, -4){10}}
\put(207,40){\line(2, 0){88}}
\put(246,18){$\nu^j$}
\put(196,65){$W_\rho^j$}
\multiput(399,100)(0,20){2}{\qbezier(0,0)(4,5)(0,10)}
\multiput(399,110)(0,20){1}{\qbezier(0,0)(-4,5)(0,10)}
\multiput(400,100)(-12,-16){4}{\qbezier(0,0)(-4,-2)(-6,-8)}
\multiput(394,92)(-12,-16){3}{\qbezier(0,0)(0,-4)(-6,-8)}
\put(356,40){\qbezier(0,0)(2,1)(1.5,3.5)}
\put(357,40){\line(-3, -4){10}}
\multiput(400,100)(12,-16){4}{\line(3, -4){8.5}}
\put(445,40){\line(3, -4){10}}
\put(357,40){\line(2, 0){88}}
\put(396,18){$\nu^j$}
\put(346,65){$W_\rho^j$}
\put(438,65){$G^j$}
\end{picture}
\begin{center}
\parbox{5.5in}{
\caption[]
{\small Contributions to $g_\mu-2$ from KK modes of the
$SU(2)_W\times U(1)_Y$ gauge fields, labeled by $c(Z^\rho)$, 
$c(A^\rho)$, $c(W^\rho)$, $c(G^\pm)$ in Eq.~(\ref{a-mu}).
\label{diagrams}}}
\end{center}
\end{figure}
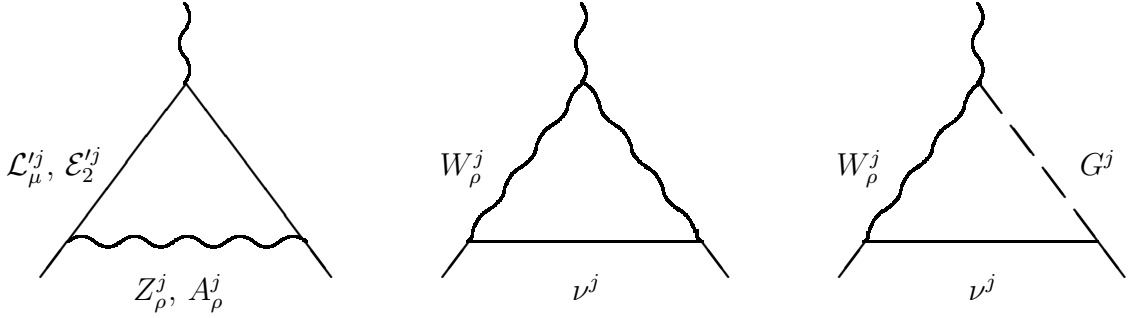

The standard model in universal extra dimensions leads to the
one-loop corrections to $a_\mu$ shown in Figs.~1 and 2. Each diagram
gives a contribution of order $(\alpha / \pi) m_\mu^2 /M_j^2$,
reflecting the decoupling of the KK modes\footnote{
The result is proportional to $(m_\mu R)^2$, unlike the linear 
dependence on $m_\mu R$ in Ref.~\cite{Casadio:2000pj},
due to the chiral couplings of the muon to the gauge boson 
KK modes.}. 
An important feature
of these diagrams is that the contributions from individual KK
levels are independent, so that the result is simply a sum over
KK levels:
\be a_\mu^{\rm KK}  = \frac{\alpha}{8\pi} \sum_{j}D_{j}
\frac{ m_\mu^2}{M_j^2} \left\{ c(Z^\rho) + c(A^\rho) + 
c(W^\rho) + c(G^\pm) + \left[ c(A^4) + 
c(Z^4) + c(W^4)\right]\delta  \right\} ~,
\label{a-mu} \ee where $D_j$ is the degeneracy of the $j$th KK
level.  The coefficients $c(Z^\rho)$, 
$c(A^\rho)$, $c(W^\rho)$ and $c(G^\pm)$ correspond to the 
diagrams shown in Fig.~1.
The coefficients $c(A^4)$, $c(Z^4)$ and $c(W^4)$, given by
diagrams which involve scalar KK modes of the 
photon, $Z$ and $W$ (see Fig.~2),  
are multiplied by the number of extra dimensions because the
higher dimensional gauge fields have $4+\delta$ components. Using
the interactions displayed in Section 2, we compute the diagrams
of Figs.~1 and 2 in the Feynman gauge, 
and obtain the following values for each of the 
coefficients:
 \bear
c(A^\rho) & = & -\frac{2}{3}\, c(A^4) \; = \;
\frac{2}{3} ~,
\\ [3.2mm]
c(Z^\rho) & = & - \frac{3 + 4\sin^2\theta_W
\cos2\theta_W}{3\sin^2 2\theta_W} ~,
 \\ [3.2mm]
c(Z^4) & = & \frac{ 1 + 12\sin^2\theta_W\cos2\theta_W}
{6\sin^2 2\theta_W}
 ~,
 \\ [3.2mm]
c(W^\rho) & = & 2 \, c(W^4) \; =  \; - c(G^\pm)  \; =   \;
-\frac{1}{3\sin^2\theta_W} ~.
\eear

\begin{figure}[t]
\centering
\begin{picture}(100,130)(160,-1)
\thicklines
\multiput(99,100)(0,20){2}{\qbezier(0,0)(4,5)(0,10)}
\multiput(99,110)(0,20){1}{\qbezier(0,0)(-4,5)(0,10)}
\put(100,100){\line(3, -4){55}}
\put(100,100){\line(-3, -4){55}}
\multiput(55,40)(16, 0){6}{\line(2, 0){10}}
\put(32,65){$\CL^{\prime j}_\mu ,\, \CE^{\prime j}_2 $}
\put(80,18){$ Z_4^j, \, A_4^j   $}
\multiput(299,100)(0,20){2}{\qbezier(0,0)(4,5)(0,10)}
\multiput(299,110)(0,20){1}{\qbezier(0,0)(-4,5)(0,10)}
\multiput(300,100)(-12,-16){4}{\line(-3, -4){8.5}}
\put(255,40){\line(-3, -4){10}}
\multiput(300,100)(12,-16){4}{\line(3, -4){8.5}}
\put(345,40){\line(3, -4){10}}
\put(255,40){\line(2, 0){90}}
\put(296,18){$\nu^j$}
\put(246,65){$W_4^j$}
\end{picture}
\begin{center}
\parbox{5.5in}{
\caption[]
{\small Contributions to $g_\mu-2$ from the
gauge fields polarized in extra dimensions, 
labeled by $c(A^4), c(Z^4), c(W^4)$ in Eq.~(\ref{a-mu}).
\label{diagrams2}}}
\end{center}
\end{figure}
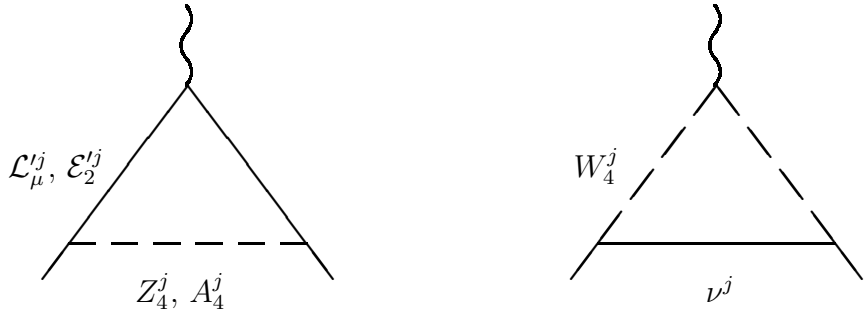

The sum of all the diagrams
takes the form \be a_\mu^{\rm KK} = \frac{\alpha}{24\pi
\sin^{2}2\theta_{W}} \left[ -3 + 4
\sin^{2}\theta_{W} - \frac{\delta}{2} \left(3 + 8
\sin^{2}\theta_{W}\right)\right]
\sum_{j}D_{j} \frac{ m_\mu^2}{M_j^2}
~, \label{a-muexplicit} \ee
and
using $\sin^{2}\theta_{W} \approx 0.231$, we find \be a_\mu^{\rm KK}
\approx - 5.8 \times 10^{-11}~( 1 + 1.2
\delta)~S_{\rm KK} ~, \label{a-mu5d} \ee
where we defined
\be
S_{\rm KK} \equiv
\sum_{j}~\frac{6D_{j}}{\pi^2}\left[\frac{ 300 \; {\rm
GeV}}{M_j}\right]^{2} ~. \label{SKK} \ee

With a single universal extra dimension, the degeneracy factor
$D_j$ is unity and the sum is convergent as in the case of 
precision electroweak observables. The smallest value for $1/R$
allowed by the electroweak data is approximately
$300$ GeV, giving $S_{\rm KK} \approx 1$, and thus leading to a
negative value for $a_\mu^{\rm KK}$ of order $10^{-10}$. This
is smaller than the final expected $1\sigma$ sensitivity of the
muon $g-2$ experiment at BNL \cite{Brown:2001mg}.
The negative sign of this small contribution increases slightly
the discrepancy between the standard model prediction
\cite{Marciano:2001qq} and experiment.

For the more interesting case of two universal extra dimensions,
the KK sum diverges logarithmically, indicating that as in the
case of the electroweak observables, important contributions can
arise from physics at scales above $M_s$ as well as below. The
contribution from physics below $M_s$ can be estimated by cutting
off the KK mode sum at an $M_j$ of order $M_s$. As noted in
Ref.~\cite{Appelquist:2000nn}, this procedure for the electroweak
observables leads to $1/R \gae 500$ GeV and $M_{s}R \lae 5$. This
leads to $S_{\rm KK} \lae 1$, and therefore $a_\mu^{\rm KK}$ is
of order $10^{-10}$ or smaller in the 6-dimensional standard
model.

\section{The 6-dimensional standard model and short distance
effects on $g_\mu-2$}
\setcounter{equation}{0}

In the previous section we showed that, with one or two universal
extra dimensions, the value of $g_\mu-2 \equiv a_\mu$ induced by
loops with standard model KK fields below the effective theory
cutoff $M_s$ is smaller than the final expected $1\sigma$
sensitivity of the muon $g-2$ experiment at BNL
\cite{Brown:2001mg}. However, physics above $M_s$ also contributes
to $a_\mu$, and these effects could be large given that the
$SU(3)_C\times SU(2)_W \times U(1)_Y$ interactions are strongly
coupled at these scales. From a low-energy effective theory point
of view, the effect of physics above $M_s$ is parametrized by
higher-dimension operators suppressed by powers of $M_s$. In the
case of one universal extra dimension, the effective
5-dimensional theory breaks down at $M_s \gae 10$ TeV, so the
operators suppressed by powers of $M_s$ are not likely to induce
a large $a_\mu$. With more dimensions the cut-off $M_s$ is
lower. We concentrate in what follows on the case of two
universal extra dimensions.

In six dimensions, the standard model is chiral as in four
dimensions and is highly constrained by anomaly cancellation and
Lorentz invariance. The quarks and leptons are 4-component Weyl
fermions of definite chirality which we label by $+$ and $-$. The
cancellation of local anomalies imposes one of the following two
chirality assignments: $\CQ_+, \CU_-, \CD_-, \CL_\mp, \CE_\pm$.
Each of these 6-dimensional chiral fermions leads in the effective
4-dimensional theory to either a left- or right-handed zero-mode
fermion depending on the orbifold boundary conditions. The
6-dimensional standard model is the only known theory that
constrains the number of fermion generations to be $n_g = 3 \;
{\rm mod} \; 3$, based on the global anomaly cancellation
condition \cite{Dobrescu:2001ae}.

The gravitational anomaly cancels only if within each generation
there is a gauge singlet fermion with 6-dimensional chirality
opposite to that of the lepton doublet \cite{Arkani-Hamed:2000hv,
Dobrescu:2001ae}. These gauge singlet fermions can have Yukawa
couplings to the Higgs and lepton doublet fields, which at one
loop give rise to a negative shift in $a_\mu$. However,
the Yukawa couplings of the zero modes have to be smaller than
$\sim 10^{-10}$ in order to avoid too large Dirac neutrino
masses. There are mechanisms to explain this small parameter,
involving additional dimensions accessible only to gravity and
the singlet fermions \cite{Arkani-Hamed:1998vp}. The large number
of singlet fermion KK modes associated with these additional
dimensions enhance the contribution to the muon anomalous
magnetic moment, but even then there is no reason to expect a
sizable $a_\mu$ from the neutrino sector.

Here we point out that the chiral 6-dimensional standard model
includes higher-dimension operators suppressed by powers of $M_s$
that can naturally have a substantial contribution to $a_\mu$. In
the 6-dimensional Lagrangian these appear as dimension-ten
operators: 
\be 
\overline{\CL}_+
\frac{i}{2}[\Gamma^\alpha,\Gamma^\beta] \frac{\hat{\lambda}_\CE
}{M_s^2} \left( C_B \frac{\hat{g}^\prime}{2} 
{\cal B}_{\alpha\beta } - C_W
\frac{\vec{\sigma}}{2} \hat{g} \vec{\cal W}_{\alpha\beta }  \right)
\CE_- {\CH} ~ + {\rm h.c.} ~, \label{HDOp}
\ee 
where
${\cal W}_{\alpha\beta }$, ${\cal B}_{\alpha\beta}$ are the 
6-dimensional  $SU(2)_W \times
U(1)_Y$ field strengths, and $C_W, C_B$ are dimensionless
parameters determined by the unknown physics above $M_s$. We have
defined them by extracting the 6-dimensional $SU(2)_W \times
U(1)_Y$ gauge couplings $\hat{g}, \hat{g}^\prime$, and charged
lepton Yukawa coupling matrix, $\hat{\lambda}_\CE$. These have inverse
mass dimension, and are related to the corresponding
4-dimensional couplings by 
\be 
\left\{ \hat{g}, \hat{g}^\prime,
\hat{\lambda}_\CE \right\} = \sqrt{2}\pi R \left\{ g, g^\prime,
\lambda_\CE \right\} ~. \ee

After the two extra dimensions are integrated out, the operator
(\ref{HDOp})
gives rise to a number of terms in the 4-dimensional Lagrangian.
Only the Higgs doublet zero-mode acquires a VEV, leading to 
interactions of the leptons with gauge bosons described by
dimension-five operators. 
Among those that involve only zero-modes, 
the following operator contributes to $a_\mu$: \be 
\frac{em_\mu}{2M_s^2}\overline{\mu}^\prime
[U^\dagger (C_B + C_W) U]_{22}
\frac{i}{2}[\gamma^\rho,\gamma^\tau] \mu^\prime F_{\rho\tau} 
~, \label{op-mmm}
\ee where $F_{\rho\tau}$ is the electromagnetic
field strength, $\mu^\prime$ is the muon mass eigenstate,
and $U$ is the unitary matrix that relates
the mass eigenstate charged leptons to the weak eigenstates. 
In general, $C_W$ and $C_B$ are $3\times 3$ matrices in flavor 
space. However, the gauge fields have generational-independent
couplings, so we expect that the flavor-dependence
of the operator (\ref{HDOp}) is due only to the presence of
the Higgs field and shows up predominantly through the
Yukawa coupling matrix $\hat{\lambda}_\CE$.
In other words, we expect $C_W$ and $C_B$ to be approximately
flavor independent:
\be
C_{W,B}^{ii^\prime} = c_{W,B} 
\left( \delta_{ii^\prime} + \epsilon_{W,B}^{ii^\prime} \right) ~,
\ee
with $\epsilon_{W,B}^{ii^\prime} \ll 1$ ($i,i^\prime =1,2,3$),
on the order of the squared lepton Yukawa couplings.
Therefore, the muon anomalous magnetic moment induced by 
the operator (\ref{op-mmm}) is given by 
 \be a_\mu^{\rm op} \approx
\frac{2m_\mu^2}{M_s^2}(c_B + c_W) ~. 
\label{formula}
\ee

If $M_s$ is taken to
be the scale where the standard model gauge interactions become
non-perturbative, then $RM_s \approx 5$. The bound on the size of
two universal extra dimensions imposed by the electroweak data is
$1/R \gae 500$ GeV \cite{Appelquist:2000nn}. 
Since anomaly cancellation in six dimensions
does not allow a straightforward supersymmetric extension 
of the standard model \cite{Dobrescu:2001ae},
the scale where the 6-dimensional standard model breaks down 
should be not much higher than a few TeV in order
to avoid fine-tuning in the Higgs sector. The result for $a_\mu^{\rm op}$
can be written as
 \be
a_\mu^{\rm op} \approx 3.6 \times 10^{-9}(c_B + c_W) 
\left(\frac{2.5 \; {\rm TeV}}{M_s}\right)^2 ~. 
\ee

The operators (\ref{HDOp}) arise at scales of order $M_s$,
and because they involve gauge fields, their
coefficients are expected to be proportional to the 
 6-dimensional gauge couplings. At the same time, these 
operators break the chiral symmetry of the leptons,
and from the Yukawa terms in the 6-dimensional Lagrangian we know
that such breaking is accompanied by Yukawa couplings.
Hence,
it is natural to expect the coefficients $C_B$ and $C_W$, defined in 
Eq.~(\ref{HDOp}) by extracting the
6-dimensional gauge and Yukawa couplings, to be of order unity at 
the scale $M_s$. 
Furthermore, upon dimensional reduction 
the volume suppresion is entirely absorbed in the
gauge and Yukawa couplings\footnote{Consequently, 
the estimate (\ref{formula}) is similar to that arising from 
higher-dimension operators associated with muon substructure
in a 4-dimensional context \cite{Lane:2001ta}.}. As a 
result, the values of 
$c_B$ and $c_W$ at scales comparable to the muon mass differ
from those at the scale $M_s$ by factors of order one,  
mostly due to the one-loop running. Notice that the theory is 
perturbative at scales below $M_s$.

We conclude that physics from above the cutoff $M_s$ of the effective,
6-dimensional standard model naturally gives a
contribution to $a_\mu$ comparable to the current sensitivity of 
the Muon $g-2$ experiment at BNL. The sign of this contribution 
cannot be determined within the effective 6-dimensional standard 
model. Future reductions in the experimental uncertainty and 
improvements in the estimate of hadronic contributions to $a_\mu$ 
would allow a measurement of $(c_B + c_W)/M_s^2$ 
in the context of universal extra dimensions.

Although the operators (\ref{HDOp}) are expected to be
approximately flavor diagonal, the constraints on flavor-changing 
neutral currents are severe enough to warrant attention.
The process $\mu^-\rightarrow e^-\gamma$ is the most constraining
in this context. The tree level decay width for this process is 
\be
\Gamma(\mu^-\rightarrow e^-\gamma) \approx
\frac{\alpha m_\mu^5}{2 M_s^4} 
\left\{[U^\dagger (C_B + C_W) U]_{12}\right\}^2 ~.
\ee
The experimental limit of $\Gamma(\mu^-\rightarrow e^-\gamma) < 3.6 \times 10^{-27}$ 
MeV \cite{Groom:2000in}
imposes a bound on the off-diagonal ($i \neq i^\prime$) elements of $C_{W,B}$,
\be
\epsilon_{W,B}^{ii^\prime} \lae 
\frac{10^{-4}}{c_{W,B}} \left(\frac{2.5 \; {\rm TeV}}{M_s}\right)^2 ~.
\ee

\section{Conclusions}
\setcounter{equation}{0}

In Ref.~\cite{Appelquist:2000nn}, it was pointed out that all the
standard model fields could propagate in a larger number of
spatial dimensions, compactified at a scale $1/R$ as small as
$300$ GeV.  In this paper, we addressed the implications of this
idea of ``universal extra dimensions" for the  muon anomalous
magnetic moment. For one or two extra
dimensions, we computed the one-loop contribution of the KK modes
of the standard model fields and found that it is too small to be
detected by the Muon ($g-2$) experiment at BNL \cite{Brown:2001mg}.
We then analyzed higher-dimension operators in the context of
the 6-dimensional standard model. For the cut-off $M_s$ in 
a range such that fine-tuning in the Higgs sector is eschewed 
(a few TeV), the contribution to the muon anomalous magnetic moment
is naturally as large as the currently quoted discrepancy
\cite{Brown:2001mg}. The sign of this contribution, however,
is determined by the unknown physics above $M_s$.

It is interesting that the natural expectation of the
contribution to $g_{\mu} - 2$ from physics above $M_s$ is 
an order of magnitude
larger than the contribution from physics below $M_s$, arising
from the KK modes of standard model fields.
This is in contrast to the case of the weak-isospin violating
$\rho$ parameter discussed in \cite{Appelquist:2000nn}. The  
dimension-ten weak-isospin violating
operator $(c_T\hat{\lambda}_\CH/2M_s^2)\left(\CH^\dagger 
D_\alpha \CH\right)^2$ has a coefficient $c_T$ (defined
by extracting the 6-dimensional quartic 
Higgs coupling $\hat{\lambda}_\CH$) of order unity 
if the weak-isospin is maximally violated by physics above $M_s$. 
The volume suppression resulting from 
integration over the two extra dimensions 
is absorbed in the Higgs and gauge couplings, 
so this operator gives  $\Delta\rho \sim M_h^2/M_s^2$ ($M_h$ is the 
Higgs boson mass), comparable to the one-loop KK contribution. 
The reason for this difference is partly that 
the one-loop contributions to $g_{\mu} - 2$ involve only lepton 
KK modes, while $\Delta\rho$ is enhanced by a color factor 
and the largeness of the top Yukawa coupling.

We emphasize that the low scale of new physics, of a few TeV, 
where the 6-dimensional standard model is expected to break down,
is an opportunity, allowing phenomenologically interesting 
higher-dimension operators, but also a challenge requiring
further study of mechanisms that suppress dangerous operators
\cite{preparation}.

\bigskip

 {\bf Acknowledgements:} \ We would like to thank William Marciano for
helpful correspondence, and we are grateful to Hsin-Chia Cheng,
Eduardo Ponton and Erich Poppitz for helpful conversations and comments.
 This work was supported by DOE under contract DE-FG02-92ER-40704.

 \vfil \end{document}